\documentclass[conference,a4paper]{IEEEtran}

\usepackage[utf8]{inputenc} 
\usepackage[T1]{fontenc}
\usepackage{url}              
\usepackage{cite}             

\usepackage[cmex10]{amsmath}  
\usepackage{mleftright}       
\mleftright                   

\usepackage{graphicx}         
\usepackage{booktabs}         

\usepackage{amssymb,amsfonts,amsthm,bm,bbm,dsfont}
\newtheorem{theorem}{Theorem}

\newtheorem{proposition}{Proposition}

\newtheorem{lemma}{Lemma}

\usepackage{soul}

\DeclareSymbolFont{bbold}{U}{bbold}{m}{n}
\DeclareSymbolFontAlphabet{\mathbbold}{bbold}

\usepackage{color}

\newcommand{\sI}{\mathcal{I}}
\newcommand{\sS}{\mathcal{S}}

\DeclareMathOperator*{\argmax}{arg\,max}
\DeclareMathOperator*{\argmin}{arg\,min}

\begin{document}

\title{Perturbation-Resilient Sets for Dynamic Service Balancing}

\author{%
  \IEEEauthorblockN{
  \textbf{Jin Sima}, \textbf{Chao Pan}  and \textbf{Olgica Milenkovic}}
   \IEEEauthorblockA{Department of Electrical and Computer Engineering, University of Illinois Urbana-Champaign, USA \\\texttt{\{jsima,chaopan2,milenkov\}@illinois.edu}
   }
}

\maketitle

\begin{abstract} 
Balanced and swap-robust minimal trades, introduced in~\cite{pan2022balanced}, are important for studying the balance and stability of server access request protocols under data popularity changes. Constructions of such trades have so far relied on paired sets obtained through iterative combining of smaller sets that have provable stability guarantees, coupled with exhaustive computer search. Currently, there exists a  nonnegligible gap between the resulting total dynamic balance discrepancy and the known theoretical lower bound. We present both new upper and lower bounds on the total service requests discrepancy under limited popularity changes. Our constructive near-optimal approach uses a new class of paired graphs whose vertices are two balanced sets with edges (arcs) that capture the balance and potential balance changes induced by limited-magnitude popularity changes (swaps). 
\end{abstract}

\section{Introduction}
\label{sec:introduction}

In real-world distributed storage systems, service (access) request control methods are used to balance out requests to servers and prevent service time bottlenecks~\cite{dau2018maxminsum,liu2019distcache,chee2020access,anderson2018service,aktacs2021service}. One approach to services balancing is to allocate different combinations of files to different servers in order to ensure both efficient content reconstruction/regeneration in the presence of disk failures and even out the ``average popularity'' of files stored on servers. To perform this allocation, a new family of Steiner systems and related combinatorial designs, termed \emph{MinMax} Steiner systems, was introduced in~\cite{dau2018maxminsum,colbourn2020popularity,colbourn2021egalitarian}. There, in addition to block-intersection constraints common to designs, one also has to impose labeling rules for the elements in a block according to popularity values in order to ensure balanced server access. 

In practice, data popularities change with time, and it is costly to redistribute files across servers after each popularity change, especially when the magnitude of changes is small. It is thus important to have combinatorial designs for which the discrepancy in the average server popularity scores is tightly restricted in the presence of small perturbations in data popularity. One possible solution is based on \emph{combinatorial trades}~\cite{hedayat1990theory,billington2003combinatorial}, mathematical objects largely unknown in the coding theory literature. As the name suggests, trades allow one to exchange collections of offending blocks, violating popularity constraints, with blocks that reduce the service request discrepancy. Trades are based on sophisticated combinatorial constructions, one of which was described in~\cite{hwang1986structure} and later adapted to the problem at hand in~\cite{pan2022balanced}.

The work~\cite{pan2022balanced} described how to construct balanced and popularity-swap-robust trades to ensure the stability of server access frequencies in the presence of limited-magnitude data popularity changes. Specifically, the authors showed that balanced and swap-robust trades of orders liner in an integer parameter $t$ can be constructed through a careful selection of so-called ``defining sets'', and that one can construct such sets with provable upper bounds on the total popularity discrepancy. Their approach relies on careful recursive groupings of small optimal defining sets and exhaustive computer search. However, due to the complex combinatorial nature of the problem, the largest instance that can be solved by exhaustive search is of size $24$ (i.e., $24$ files of different popularity scores), which leads to a nonnegligible gap between the theoretical lower and upper bounds on popularity discrepancy.

The contributions of this paper are two-fold. First, we present a recursive construction (Section~\ref{sec:upper_bound}) that can achieve a significantly smaller total discrepancy under magnitude-one popularity changes, $\frac{8}{5}(t-\frac{1}{4})$, compared to the currently best reported discrepancy of $2t+O(1)$. Second, we establish a significantly tighter lower bound $\frac{3}{2}(t-\frac{2}{3})$ (Section~\ref{sec:lower_bound}) on the total set discrepancy when compared to the lower bound $\frac{2}{3}(t-\frac{1}{3})$ proved in~\cite{pan2022balanced}. Our proof techniques use new families of paired graphs that simultaneously describe the balance and discrepancy of servers as well as the potential disbalance increase induced by popularity transpositions.

\section{Preliminaries}
\label{section:preliminary}
Given the space restrictions, we do not provide details regarding the construction of access-balancing protocols and instead refer the interested reader to~\cite{pan2022balanced}. Instead, we focus on optimizing the building blocks of swap-robust trades, termed \emph{companion sets}.  To this end, assume that we are given a partition of the set of integers $T=[1,4t]=\{1,\ldots,4t\}$, corresponding to popularity rankings of $4t$ files, into $2t$ (disjoint) sets $(S_1,\ldots,S_{2t})$, that $\forall i\in [t]$ it satisfies that
$$
|S_{2i-1}|=|S_{2i}|=2,\; \Sigma(S_{2i-1})=\Sigma(S_{2i}).
$$
Here we use $\Sigma(S_{2i-1})=\sum_{j\in S_{2i-1}} j$ to denote the sum of elements within the set $S_{2i-1}$. The pairs $(S_{2i-1}, S_{2i})$ are referred to as \emph{balanced companion sets}, while the set of all balanced companion sets $(S_1,\ldots, S_{2t})$ is referred to as a \emph{balanced defining set}~\cite{pan2022balanced}. If the sum-constraint is not satisfied, we say that the companion (and corresponding defining) sets are imbalanced. The total imbalance is measured by the total \emph{discrepancy}, defined as
$\sum_{i=1}^{t}|\Sigma(S_{2i-1})-\Sigma(S_{2i})|.$ There exist a super-exponential number of valid balanced defining sets, and the key problem is to find balanced companion sets that have the smallest worst-case total set discrepancy in the presence of limited-magnitude data popularity changes (to be formally described later). Changes in popularity change the labels of elements in companion sets and consequently the sum of their elements. As a result, upon relabeling induced by popularity changes, the new resulting pairs $(S_{2i-1}^\prime, S_{2i}^\prime)$ may no longer have the same sum, leading to access imbalance. 

For simplicity, we only consider data popularity change of magnitude $1$, meaning that we consider \emph{adjacent} swaps of labels or popularity values $I=\{(i_1,j_1),(i_2,j_2),\ldots,(i_m,j_m)\},$ where $|i_\ell-j_\ell|=1$ and $i_1,\ldots,i_m,j_1,\ldots,j_m$ are all distinct. As an example, for $t=2$ and $T=[1,8]$, the allowed popularity swaps are subsets of $\{(1,2),(2,3),(3,4),(4,5),(5,6),(6,7),(7,8)\}$ such that the same element is not included in two different swaps, as swaps with a common element correspond to \emph{nonadjacent} swaps and hence larger popularity change magnitudes. As an example, $\{{(1,2),(3,4),(5,6)\}}$ is allowed, while $\{{(1,2),(2,3),(5,6),(7,8)\}}$ is not. 

In the absence of popularity swaps, the total set discrepancy of balanced defining sets equals $0$. The \emph{worst-case total balance discrepancy for a given defining set} is the largest total set discrepancy that any valid collection of swaps $I$ could possibly induce. The goal of our work is to find (near) optimal defining sets that have the smallest worst-case total set discrepancy. For example, when $t=2$, the companion sets 
$$
S_1=\{1,8\},S_2=\{3,6\},S_3=\{2,7\},S_4=\{4,5\},
$$
constitute an example of an optimal defining set. The swaps $\{(1,2),(5,6)\}$ lead to changes in the companion sets 
$$
S_1^\prime=\{2,8\},S_2^\prime=\{3,5\},S_3^\prime=\{1,7\},S_4^\prime=\{4,6\},
$$
and a total set discrepancy of $|(2+8)-(3+5)|+|(1+7)-(4+6)|=4$, which is the worst-case total set discrepancy for this case. For other choices of defining sets, such as $S_1=\{1,4\},S_2=\{2,3\},S_3=\{5,8\},S_4=\{6,7\}$, the worst-case total set discrepancy is $\geq 4$ ($6$ for the given example), making the defining sets suboptimal.

Formally, if we denote the set of all allowed collections of adjacent swaps for $4t$ elements by $\sI_t$, the set of all valid collections of defining sets by $\sS_t$, then the goal is to find a defining set $(S_1^*,\ldots,S_{2t}^*)$ such that its worst-case total discrepancy $\max_{I_t\in\sI_t} D(S_1^*,\ldots,S_{2t}^*;I_t;t)$ is the smallest possible, i.e.,
\begin{align}
&(S_1^*,\ldots,S_{2t}^*)=\argmin_{(S_1,\ldots,S_{2t})\in\mathcal{S}_t}\max_{I_t\in\mathcal{I}_t}D(S_1,\ldots,S_{2t};I_t;t), \\
&D^*(t)=\min_{(S_1,\ldots,S_{2t})\in\mathcal{S}_t}\max_{I_t\in\mathcal{I}_t}D(S_1,\ldots,S_{2t};I_t;t),\\
&D(S_1,\ldots,S_{2t};I_t;t)=\sum^{t}_{i=1}\left|\Sigma(S_{2i-1}^\prime)-\Sigma(S_{2i}^\prime)\right|,
\end{align}
where $(S_1^\prime,\ldots,S_{2t}^\prime)$ denote the sets after popularity swaps.

Note that it is computationally hard to find $(S_1^*,\ldots,S_{2t}^*)$ by exhaustive search when $t$ is even moderately large, as the size of search space grows super-exponentially. A recursive construction was proposed in~\cite{pan2022balanced}, for which the building blocks are optimal defining sets for small instances of $t$ (i.e., $t\in\{1,2,3,4,5,6\}$) that can be found via computer search. The authors also derived a lower bound on the worst-case total discrepancy of the form $\frac{2}{3}(t-\frac{1}{3})$. 

In what follows, we improve this lower bound to $\frac{3t-2}{2}$, and provide a completely analytic recursive approach that provably achieves a smaller worst-case total set discrepancy compared to the previously reported one. Our results are summarized in the next theorem.

\begin{theorem}
For any integer $t>0$, we have
\begin{align}\label{eq:lower}
   D^*(t)\ge \frac{3t-2}{2}. 
\end{align}
Moreover, for any positive integer $z\ge 2$ and $t = 5 \cdot 2^{z-2}-1$, we have
\begin{align}\label{eq:upper}
   D^*(t)\le 2^{z+1}-2,
\end{align}
which implies $D^*(t)\le \frac{8t-2}{5}$. The upper and lower bound only differ by a constant factor $1.07$.
\end{theorem}

\section{The Recursive Construction}
\label{sec:upper_bound}
\begin{figure}[t]
\centering
\includegraphics[width=1\linewidth]{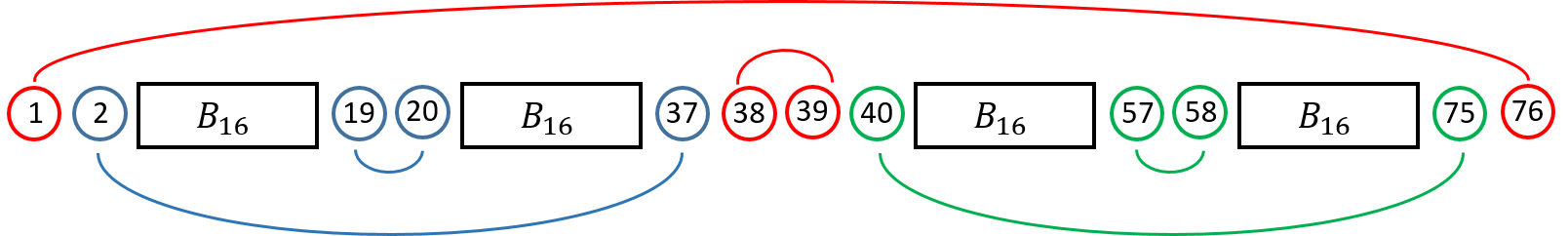}
\caption{An example of the recursive construction for $z=4,t=19$. We use the companion sets in~\eqref{eq:comp16}, denoted by $B_{16}$, as building blocks. Integers within the same companion sets have the same color.}
\label{fig:examplepartition}
\end{figure}

In what follows, we provide constructions for defining sets $(S_1,\ldots,S_{2t})\in\mathcal{S}_t$ for values of $t$ that satisfy $t=5\cdot2^{z-2}-1$, $z\ge 2$, such that the total discrepancy is upper bounded by
$$
\max_{I_t\in\sI_t} D(S_1,\ldots,S_{2t};I_t;t) \leq \frac{8t}{5}-\frac{2}{5}.
$$
To begin with, consider the case $z=2$ and $t=5\cdot 2^{z-2}-1=4$ and add superscripts to sets $S$ to indicate the value of $z$. Using exhaustive search we can show that the unique optimal defining set choice for $t=4$ equals
\begin{align}\label{eq:comp16}
&S^2_1=\{1,16\},S^2_2=\{8,9\},S^2_3=\{2,7\},S^2_4=\{4,5\},\\
&S^2_5=\{10,15\},S^2_6=\{12,13\},S^2_7=\{3,14\},S^2_8=\{6,11\}.\nonumber
\end{align}
Its discrepancy (upon performing the worst-case swaps) equals $6=2^{z+1}-2$. Next, we describe a recursive construct for defining sets $(S^{z+1}_1,\ldots,S^{z+1}_{2t_1})$ for $t_1=5\cdot 2^{(z+1)-2}-1$, based on $(S^{z}_1,\ldots,S^{z}_{2t_2})$ for $t_2=5\cdot 2^{z-2}-1$:
\begin{align*}
    &S^{z+1}_{i}=S^{z}_{i}+1, \forall i\in[5\cdot 2^{z-1}-1],\\
    &S^{z+1}_{i}=S^{z}_{i-5\cdot 2^{z-1}+2}+5 \cdot 2^{z}-1,
    \forall i\in[5 \cdot 2^{z-1}-1,5 \cdot 2^{z}-4],\\
    &S^{z+1}_{5 \cdot 2^{z}-3}=\{1,5 \cdot 2^{z+1}-4\},\\
    &S^{z+1}_{5 \cdot 2^{z}-2}=\{5 \cdot 2^{z}-2,5 \cdot 2^{z}-1\},
\end{align*}
where for an integer set $S$ and an integer $a$, we define $S+a=\{x+a:x\in S\}$. An example of the construction for $z=4$ is depicted in Fig.~\ref{fig:examplepartition}. The intuition behind the construction is as follows: once we fix the last companion sets to $S^{z+1}_{5 \cdot 2^{z}-3}=\{1,5\cdot2^{z+1}-4\}$ and $S^{z+1}_{5 \cdot 2^{z}-2}=\{5 \cdot 2^{z}-2,5 \cdot 2^{z}-1\}$, the elements used in the other $t_1-1$ companion sets must come from two disjoint and symmetric intervals $[2,5\cdot 2^{z}-3]$ and $[5\cdot 2^{z},5 \cdot 2^{z+1}-5]$. Therefore, we can reuse the construction of companion sets $(S^{z}_1,\ldots,S^{z}_{2t_2})$ for $t_2=5 \cdot2^{z-2}-1$. 

Next, we upper-bound the worst-case total discrepancy for the recursively constructed collection of companion sets as follows. Let 
$$
d_z=\max_{I_t\in\mathcal{I}_t}D(S^z_1,\ldots,S^z_{2t};I_t;t),\; t=5 \cdot 2^{z-2}-1.
$$ 
We have the following lemma.

\begin{lemma}\label{lemma:dzrecursion}
For $z\ge 2$, we have $d_{z+1}\le 2d_z+2$.
\end{lemma}

\begin{IEEEproof}
The proof relies on examining scenarios for the discrepancy change induced by an allowed set of swaps $I_t$:
\begin{enumerate}
    \item $|\Sigma(S^{\prime z+1}_{5\cdot 2^{z}-3})-\Sigma(S^{\prime z+1}_{5 \cdot 2^{z}-2})|=0$.
    \item $|\Sigma(S^{\prime z+1}_{5 \cdot 2^{z}-3})-\Sigma(S^{\prime z+1}_{5 \cdot 2^{z}-2})|=1$.
    \item $|\Sigma(S^{\prime z+1}_{5 \cdot 2^{z}-3})-\Sigma(S^{\prime z+1}_{5 \cdot 2^{z}-2})|=2$.
\end{enumerate}

\textbf{Case 1):} In this setting we have 
\begin{align*}
&|I_t\cap\{(1,2),(5 \cdot 2^{z}-3,5 \cdot 2^{z}-2)\}|\\
=&|I_t\cap\{(5 \cdot 2^{z}-1,5 \cdot 2^{z}),(5 \cdot 2^{z+1}-3,5 \cdot 2^{z+1}-4)\}|. 
\end{align*}

If $|I_t\cap\{(1,2),(5\cdot 2^{z}-3,5 \cdot 2^{z}-2)\}|=0$, then the discrepancy corresponding to the sets $(S^{\prime z+1}_{1},\ldots,S^{\prime z+1}_{5\cdot 2^{z-1}-2})$, as well as to the sets  $(S^{\prime z+1}_{5 \cdot 2^{z-1}-1},\ldots,S^{\prime z+1}_{5 \cdot 2^{z}-4})$ is at most $d_z$. Thus, the total discrepancy is at most $2d_z$. 

If $|I_t\cap\{(1,2),(5 \cdot 2^{z}-3,5 \cdot 2^{z}-2)\}|=1$, then the discrepancy corresponding to the sets $(S^{\prime z+1}_{1},\ldots,S^{\prime z+1}_{5 \cdot 2^{z-1}-2})$, as well as to the sets $(S^{\prime z+1}_{5\cdot 2^{z-1}-1},\ldots,S^{\prime z+1}_{5 \cdot 2^{z}-4})$ is at most $d_z+1$. Hence, the total discrepancy is at most $2d_z+2$. 

Similarly, if $|I_t\cap\{(1,2),(5 \cdot 2^{z}-3,5 \cdot 2^{z}-2)\}|=2$, then the set discrepancies corresponding to $(S^{\prime z+1}_{1},\ldots,S^{\prime z+1}_{5\cdot 2^{z-1}-2})$ and to $(S^{\prime z+1}_{5 \cdot 2^{z-1}-1},\ldots,S^{\prime z+1}_{5 \cdot 2^{z}-4})$ are at most $d_z$.
Then, 
the total discrepancy of $(S^{\prime z+1}_{1},\ldots,S^{\prime z+1}_{5 \cdot 2^{z}-2})$ is at most $2d_z$.

\textbf{Case 2):} For this scenario, either one of the two conditions holds:
\begin{align*}
&|I_t\cap\{(1,2),(5 \cdot 2^{z}-3,5 \cdot 2^{z}-2)\}|=1, \\ 
&|I_t\cap\{(5 \cdot 2^{z}-1,5 \cdot 2^{z}),(5 \cdot 2^{z+1}-3,5 \cdot 2^{z+1}-4)\}|=1.
\end{align*}
By symmetry, we can assume that $|I_t\cap\{(1,2),(5 \cdot 2^{z}-3,5 \cdot 2^{z}-2)\}|=1$. Then either one of the following holds:
\begin{align*}
&|I_t\cap\{(5 \cdot 2^{z}-1,5 \cdot 2^{z}),(5 \cdot 2^{z+1}-3,5 \cdot 2^{z+1}-4)\}|=0, \\ 
&|I_t\cap\{(5 \cdot 2^{z}-1,5 \cdot 2^{z}),(5 \cdot 2^{z+1}-3,5 \cdot 2^{z+1}-4)\}|=2.
\end{align*}
Hence, the discrepancy induced by the sets $(S^{\prime z+1}_{1},\ldots,S^{\prime z+1}_{5 \cdot 2^{z-1}-2})$ is at most $d_z+1$ and the discrepancy induced by $(S^{\prime z+1}_{5 \cdot 2^{z-1}-1},\ldots,S^{\prime z+1}_{5 \cdot 2^{z}-4})$ is at most $d_z$. Hence, the total discrepancy for all sets $(S^{\prime z+1}_{1},\ldots,S^{\prime z+1}_{5 \cdot 2^{z}-2})$ is at most $d_z+1+d_z+1=2d_z+2$ where the extra $1$ comes from the discrepancy of $(S^{\prime z+1}_{5\cdot 2^{z}-3},S^{\prime z+1}_{5\cdot 2^{z}-2})$.

\textbf{Case 3):} Either one of the following holds:
\begin{align*}
&|I_t\cap\{(1,2),(5 \cdot 2^{z}-3,5 \cdot 2^{z}-2)\}|=2, \\
&|I_t\cap\{(5 \cdot 2^{z}-1,5 \cdot 2^{z}),(5 \cdot 2^{z+1}-3,5 \cdot 2^{z+1}-4)\}|=2.
\end{align*}
Similarly to Case 2), by symmetry we can assume that $|I_t\cap\{(1,2),(5 \cdot 2^{z}-3,5 \cdot 2^{z}-2)\}|=2$ and thus $|I_t\cap\{(5 \cdot 2^{z}-1,5 \cdot 2^{z}),(5 \cdot 2^{z+1}-3,5 \cdot 2^{z+1}-4)\}|=0$. Then, the discrepancy induced by $(S^{\prime z+1}_{1},\ldots,S^{\prime z+1}_{5 \cdot 2^{z-1}-2})$ is at most $d_z$, while the discrepancy induced by $(S^{\prime z+1}_{5 \cdot 2^{z-1}-1},\ldots,S^{\prime z+1}_{5 \cdot 2^{z}-4})$ is at most $d_z$. Hence, the total discrepancy of all sets $(S^{\prime z+1}_{1},\ldots,S^{\prime z+1}_{5 \cdot 2^{z}-2})$ is at most $d_z+d_z+2=2d_z+2$, where the extra $2$ comes from the discrepancy of $(S^{\prime z+1}_{5\cdot 2^{z}-3},S^{\prime z+1}_{5\cdot 2^{z}-2})$.
\end{IEEEproof}
By Lemma~\ref{lemma:dzrecursion} and the fact that $d_2=6$, we arrive at
\begin{align}\label{eq:recur_ub}
d_z\le 2^{z+1}-2=\frac{8t}{5}-\frac{2}{5}.
\end{align}

\section{A Lower Bound on $D^*(t)$}
\label{sec:lower_bound}
We now prove~\eqref{eq:lower}. Before that, we present the definitions needed for our proof. 

\begin{figure}[t]
\centering
\includegraphics[width=0.9\linewidth]{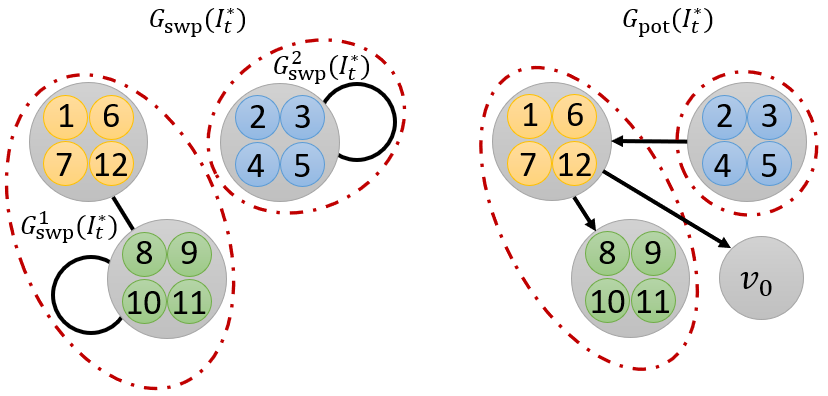}
\caption{Two auxiliary graphs $G_{\text{\text{swp}}}(I_t^*)$ and $G_{\text{\text{pot}}}(I_t^*)$ defined for the worst-case popularity swaps $I_t^*=\{(2,3),(8,9),(11,12)\}$ and $t=3$. Each gray circle, containing $4$ elements, represents a pair of companion sets and represents one node in both graphs. Two connected components $G_{\text{\text{swp}}}^1(I_t^*),G_{\text{\text{swp}}}^2(I_t^*)$ are circled out in $G_{\text{\text{swp}}}(I_t^*)$. For analytical convenience, the compositions of connected components are always $G_{\text{\text{swp}}}^1(I_t^*),G_{\text{\text{swp}}}^2(I_t^*)$, no matter if the underlying graph is $G_{\text{\text{swp}}}(I_t^*)$ or $G_{\text{\text{pot}}}(I_t^*)$.}
\label{fig:graph_example}
\end{figure}

Fix the companion sets $(S_1,\ldots,S_{2t})\in \mathcal{S}_t$. For any collection of allowed swaps $I_t\in\mathcal{I}_t$, define an unweighted, undirected graph $G_{\text{swp}}(I_t)=(V_{\text{swp}}(I_t),E_{\text{swp}}(I_t))$ that describes the swap set $I_t$. Specifically, the node set $V_{\text{swp}}(I_t)=\{v_i\}^t_{i=1},$ where the node $v_i$ corresponds to the companion sets $(S_{2i-1},S_{2i})$. For any two nodes $v_{i_1},v_{i_2}\in V_{\text{swp}}(I_t)$, there exists an edge between $v_{i_1}$ and $v_{i_2}$ if there exists a swap $(i,i+1)\in I_t$ such that $i$ and $i+1$ are in the sets $S_{2i_1-1}\cup S_{2i_1}$ and $S_{2i_2-1}\cup S_{2i_2}$, respectively. Note that $i_1$ and $i_2$ can be the same, meaning that $G_{\text{swp}}(I_t)$ is allowed to have self loops. In addition, multiple edges are also allowed between the same pair of nodes. One example of $G_{\text{swp}}(I_t)$ for $t=3$ and $I_t=\{(2,3),(8,9),(11,12)\}$ is shown in Fig.~\ref{fig:graph_example}. Note that a similar (yet different) definition to that of $G_{\text{swp}}(I_t)$ was also used in~\cite{pan2022balanced}.

In addition to $G_{\text{swp}}(I_t)$, we also define a directed graph $G_{\text{pot}}(I_t)=(V_{\text{pot}}(I_t),E_{\text{pot}}(I_t))$ that describes the \text{pot}ential discrepancy change induced by swaps $(i,i+1)$ that are not included in $I_t$. The node set $V_{\text{pot}}(I_t)=V_{\text{swp}}(I_t)\cup\{v_0\}$ equals the node set $V_{\text{swp}}(I_t)$ augmented by an auxiliary node $v_0$ that acts as a virtual companion set and is discussed later. To define the directed edge set (i.e., arc set) $E_{\text{pot}}(I_t)$, we consider all potential swaps $(i,i+1)\notin I_t$ for $i\in [4t-1]$. Note that the potential swap $(i,i+1)\notin I_t$ and the selected swaps in $I_t$ are allowed to share an element. For a swap $(i,i+1)\notin I_t$, an arc  $(v_{i_1},v_{i_2})$ directed from $v_{i_1}$ to $v_{i_2}$ exists if  
$(i,i+1)$ satisfies one of the following: 
\begin{enumerate}
    \item $i\in S_{2i_1}$, $i+1\in (S_{2i_2-1}\cup S_{2i_2})\backslash S_{2i_1}$, and $\Sigma(S_{2i_1-1}^\prime)<\Sigma(S_{2i_1}^\prime)$ after performing the swaps in $I_t$.
    \item $i+1\in S_{2i_1}$, $i\in (S_{2i_2-1}\cup S_{2i_2})\backslash S_{2i_1}$, and $\Sigma(S_{2i_1-1}^\prime)>\Sigma(S_{2i_1}^\prime)$ after performing the swaps in $I_t$.
    \item $i\in S_{2i_1-1}$, $i+1\in (S_{2i_2-1}\cup S_{2i_2})\backslash S_{2i_1-1}$, and $\Sigma(S_{2i_1-1}^\prime)>\Sigma(S_{2i_1}^\prime)$ after performing the swaps in $I_t$.
    \item $i+1\in S_{2i_1-1}$, $i\in (S_{2i_2-1}\cup S_{2i_2})\backslash S_{2i_1-1}$, and $\Sigma(S_{2i_1-1}^\prime)<\Sigma(S_{2i_1}^\prime)$ after performing the swaps in $I_t$.
    \item $i\in S_{2i_1-1}$, $i+1\in (S_{2i_2-1}\cup S_{2i_2})\backslash S_{2i_1-1}$, and $\Sigma(S_{2i_1-1}^\prime)=\Sigma(S_{2i_1}^\prime)$ after performing the swaps in $I_t$.
    \item $i+1\in S_{2i_1}$, $i\in (S_{2i_2-1}\cup S_{2i_2})\backslash S_{2i_1}$, and $\Sigma(S_{2i_1-1}^\prime)=\Sigma(S_{2i_1}^\prime)$ after performing the swaps in $I_t$.
\end{enumerate}
Intuitively, $(v_{i_1},v_{i_2})\in E_{\text{pot}}(I_t)$ if $i$ and $i+1$ are in $S_{2i_1-1}\cup S_{2i_1}$ and $S_{2i_2-1}\cup S_{2i_2}$, respectively, and the swap $(i,i+1)$ further increases the discrepancy $|\Sigma(S_{2i_1-1}^\prime)-\Sigma(S_{2i_1}^\prime)|$ of the companion sets $(S_{2i_1-1}^\prime,S_{2i_1}^\prime)$, when this discrepancy is nonzero. In addition, if the discrepancy of the companion sets $(S_{2i_1-1}^\prime,S_{2i_1}^\prime)$ is $0$, we have $(v_{i_1},v_{i_2})\in E_{\text{pot}}(I_t)$ only when the swap $(i,i+1)$ can results in positive set difference $\Sigma(S_{2i_1-1}^\prime)-\Sigma(S_{2i_1}^\prime)$. Note that conditions $5)$ and $6)$ are introduced to guarantee that the swaps corresponding to multiple arcs can simultaneously increase the set discrepancy of the same pair of companion sets. Again, $i_1=i_2$, i.e., self-loops, are allowed in $G_{\text{pot}}(I_t)$. Moreover, multiple arcs $(v_{i_1},v_{i_2})\in E_{\text{pot}}$ are allowed for the same node pair $(v_{i_1},v_{i_2})$. 

In addition to arcs among nodes $\{v_i\}^t_{i=1}$ in $G_{\text{pot}}(I_t)$, we add an arc $(v_{i_1},v_0)\in E_{\text{pot}}(I_t)$ if one of the following holds:
\begin{enumerate}
    \item $1\in S_{2i_1-1}\cup S_{2i_1}$ and the swap $(0,1)$ can increase the discrepancy of the companion sets $(S_{2i_1-1}^\prime,S_{2i_1}^\prime)$ when $\Sigma(S_{2i_1-1}^\prime)\ne\Sigma(S_{2i_1}^\prime)$;
    \item $1\in S_{2i_1}$ when $\Sigma(S_{2i_1-1}^\prime)=\Sigma( S_{2i_1}^\prime)$.
\end{enumerate}
Note that the swap $(0,1)$ is not allowed in $I_t$ and is only included in $E_{\text{pot}}(I_t)$ for the purposes of simplified analysis. Similarly, we define an arc $(v_{i_2},v_0)\in E_{\text{pot}}(I_t)$ for the swap $(4t,4t+1)$. One example of $G_{\text{pot}}(I_t)$ when $t=3$ and $I_t=\{(2,3),(8,9),(11,12)\}$ is shown in Fig.~\ref{fig:graph_example}.

Given defining sets $(S_1,$ $\ldots, S_{2t})$, let $I_t^*$ be a swap-set of smallest size among all swap sets that lead to the worst-case total discrepancy, i.e.,
$$
I_t^*\in \argmax_{I_t\in\mathcal{I}_t} D(S_1,\ldots,S_{2t};I_t;t).
$$
We show next that
\begin{align}\label{eq:discrepancy}
    D(S_1,\ldots, S_{2t};I_t^*;t)=2|I_t^*|,
\end{align}
implying that each pair $(i,i+1)\in I_t^*$ increases the total discrepancy by $2$. This holds because a swap $(i,i+1)$ either increases or decreases the discrepancy of two companion sets $(S_{2i_1-1},S_{2i_1})$ and $(S_{2i_2-1},S_{2i_2})$ that contain $i$ and $i+1$ by $1$, unless $i$ and $i+1$ are in the same set $S_{2i_1-1}$ or $S_{2i_1}$, which is a case that can be ignored. Therefore, the contribution of the swap $(i,i+1)\in I_t^*$ to the total discrepancy is $2$, $0$, or $-2$. Since $I_t^*$ is of the smallest size, each swap in $I_t^*$ contributes $2$ to the total set discrepancy. Hence,~\eqref{eq:discrepancy} holds.

Since our subsequent analysis does not depend on the value of $t$, we henceforth omit the subscript $t$ from the notation. For simplicity, we also write $G_{\text{swp}}=G_{\text{swp}}(I^*)=(V_{\text{swp}},E_{\text{swp}})$, and $G_{\text{pot}}=G_{\text{pot}}(I^*)=(V_{\text{pot}},E_{\text{pot}})$. The graph $G_{\text{swp}}$ can be partitioned into a set of connected components $G^j_{\text{swp}}=(V^j_{\text{swp}},E^j_{\text{swp}})$, $j\in[J]$, as illustrated by the example in Fig.~\ref{fig:graph_example}.
For $v_i\in V_{\text{swp}}\cup\{v_0\}$, let 
$$
d^{\text{pot}}_{\text{in}}(v_i)=|\{(v_j,v_i):(v_j,v_i)\in E_{\text{pot}},v_j\in V_{\text{swp}}\backslash\{v_i\}\}|
$$
be the in-degree of node $v_i$ and 
$$
d^{\text{pot}}_{\text{out}}(v_i)=|\{(v_i,v_j):(v_i,v_j)\in E_{\text{pot}},v_j\in V_{\text{swp}}\backslash\{v_i\}\}|
$$
be the out-degree of node $v_i$ in $G_{\text{pot}}$. Note that except for the auxiliary node $v_0$, it is assumed that $V_{\text{swp}}$ and $V_{\text{pot}}$ share the same vertex set $\{v_i\}^t_{i=1}$. In the following, we interchangeably use $v_i$ to denote a node in $V_{\text{swp}}$ or a node in $V_{\text{pot}}$. In addition, for an arbitrary vertex subset $V_{\subseteq} \subseteq \{v_i\}^t_{i=0}$, we use
$$
d^{\text{pot}}_{\text{in}}(V_{\subseteq})=\sum_{v\in V_{\subseteq}}d^{\text{pot}}_{\text{in}}(v),
d^{\text{pot}}_{\text{out}}(V_{\subseteq})=\sum_{v\in V_{\subseteq}}d^{\text{pot}}_{\text{out}}(v).
$$ 

\begin{lemma}\label{lemma:vertexdegreeineq}
For all $j\in [J]$, where $J$ stands for the number of connected components, we have
\begin{align}\label{eq:vertexineq}
     d^{\text{pot}}_{\text{in}}(V^j_{\text{swp}})-d^{\text{pot}}_{\text{out}}(V^j_{\text{swp}})\le |V^j_{\text{swp}}|+4(|E^j_{\text{swp}}|-|V^j_{\text{swp}}|).
 \end{align}
\end{lemma}
Note that $d^{\text{pot}}_{\text{in}}(V_{\text{swp}})-d^{\text{pot}}_{\text{out}}(V_{\text{swp}})\ge 0-d^{\text{pot}}_{\text{in}}(v_0)\ge -2$, since $d^{\text{pot}}_{\text{out}}(v_0)=0$ and $d^{\text{pot}}_{\text{in}}(V_{\text{swp}}\cup\{v_0\})-d^{\text{pot}}_{\text{out}}(V_{\text{swp}}\cup\{v_0\})=0$.  
Hence, we have
\begin{align*}
    -2\le |V_{\text{swp}}|+4(|E_{\text{swp}}|-|V_{\text{swp}}|),
\end{align*}
which implies that 
\begin{align}\label{eq:mainineq}
    2|E_{\text{swp}}|\ge \frac{3|V_{\text{swp}}|}{2}-1.
\end{align}
Since $E_{\text{swp}}=|I^*|$ by the definition of $G_{\text{swp}}$, ~\eqref{eq:discrepancy} and~\eqref{eq:mainineq} imply 
$$
\max_{I\in\mathcal{I}}D(S_1,\ldots,S_{2t};I)\ge \frac{3t-2}{2}
$$ 
for any $(S_1,\ldots,S_{2t})$. Thus we have~\eqref{eq:lower}.

We now prove Lemma \ref{lemma:vertexdegreeineq}. 
Note that for any $j\in J$, we have
\begin{align*}
&d^{\text{pot}}_{\text{in}}(V^j_{\text{swp}})-d^{\text{pot}}_{\text{out}}(V^j_{\text{swp}})\\
=&|\{(v',v):(v',v)\in E_{\text{pot}},v'\notin V^j_{\text{swp}},v\in V^j_{\text{swp}}\}|\\
&-|\{(v,v'):(v,v')\in E_{\text{pot}},v'\notin V^j_{\text{swp}},v\in V^j_{\text{swp}}\}|, 
\end{align*}
i.e., $d^{\text{pot}}_{\text{in}}(V^j_{\text{swp}})-d^{\text{pot}}_{\text{out}}(V^j_{\text{swp}})$ equals the number of arcs from nodes outside $V^j_{\text{swp}}$ to nodes in $V^j_{\text{swp}}$, minus the number of arcs from nodes inside $V^j_{\text{swp}}$ to nodes outside $V^j_{\text{swp}}$. Henceforth, for any subset $V_{\subseteq}\subseteq V_{\text{swp}}\cup\{v_0\}$, let
$$
\text{in}(V_{\subseteq})=|\{(v',v):(v',v)\in E_{\text{pot}},v'\notin V_{\subseteq},v\in V_{\subseteq}\}|
$$
be the number of ingoing arcs in $E_{\text{pot}}$ that emanate from nodes outside $V_{\subseteq}$ to nodes in $V_{\subseteq}$, and let
$$
\text{out}(V_{\subseteq})=|\{(v,v'):(v,v')\in E_{\text{pot}},v'\notin V_{\subseteq},v\in V_{\subseteq}\}|
$$ 
be the number of outgoing arcs from nodes in $V_{\subseteq}$ to nodes outside $V_{\subseteq}$, respectively. Then,~\eqref{eq:vertexineq} is equivalent to
\begin{align}\label{eq:equivvertexineq}
    \text{in}(V^j_{\text{swp}}) - \text{out}(V^j_{\text{swp}})\le |V^j_{\text{swp}}|+4(|E^j_{\text{swp}}|-|V^j_{\text{swp}}|).
\end{align}
We now prove~\eqref{eq:equivvertexineq}.  
For any vertex set $V_{\subseteq} \subseteq V_{\text{swp}}$, let
$$
d(V_{\subseteq})=|\{\{v,v'\}:\{v,v'\}\in E_{\text{swp}},\{v,v'\}\cap V_{\subseteq}\ne\varnothing\}|
$$
denote the number of edges in $E_{\text{swp}}$ that are incident to at least one node in $V_{\subseteq}$. 

\begin{proposition}\label{prop:removeandadd}
 For any subset $V_{\subseteq} \subseteq V_{\text{swp}}$, $\text{in}(V_{\subseteq})\le d(V_{\subseteq})$. 
\end{proposition}
\begin{IEEEproof}
Note that every edge in $E_{\text{swp}}$ corresponds to a swap in $I^*$ and every arc in $E_{\text{pot}}$ corresponds to a swap not in $I^*$. 
Suppose to the contrary that $\text{in}(V_{\subseteq})> d(V_{\subseteq})$, and remove all swaps $(i,i+1)\in I^*$ that correspond to edges in 
$$
E^{\text{swp}}(V_{\subseteq})=\{\{v,v'\}:\{v,v'\}\in E_{\text{swp}},\{v,v'\}\cap V_{\subseteq}\ne\varnothing\},
$$
which are edges incident to at least one node in $V_{\subseteq}$. Then, we add swaps that correspond to ingoing arcs in 
$$
E^{\text{pot}}(V_{\subseteq})=\{(v',v):(v',v)\in E_{\text{pot}},v'\notin V_{\subseteq},v\in V_{\subseteq}\}
$$ 
that emanate from nodes outside $V_{\subseteq}$ to nodes in $V_{\subseteq}$. We show next that each added swap contributes $2$ to the total discrepancy. Note that by the definition of an arc in $E_{\text{pot}}$, every swap $(i,i+1)$ that corresponds to an arc $(v_{k_2},v_{k_1})\in E^{\text{pot}}(V_{\subseteq})$ increases the discrepancy of the companion set $(S_{2k_2-1}^\prime,S_{2k_2}^\prime)$. It therefore suffices to show that the same swap $(i,i+1)\in E^{\text{pot}}(V_{\subseteq})$ contributes $+1$ to the set discrepancy of $(S_{2k_1-1}^\prime,S_{2k_1}^\prime)$ after removing the swaps corresponding to the edges in $E^{\text{swp}}(V_{\subseteq})$, for any $v_{k_1}\in V_{\subseteq}$.  
Otherwise, there would be two arcs $(v_{k_2},v_{k_1}),(v_{k_3},v_{k_1})\in E^{\text{pot}}(V_{\subseteq})$ ($k_2$ and $k_3$ can be equal), such that their corresponding swaps $(i,i+1)$ and $(j,j+1)$ contribute $+1$ and $-1$ to the set discrepancy of $(S_{2k_1-1}^\prime, S_{2k_1}^\prime)$, respectively, after swap removal and addition. This implies that the swaps $(i,i+1)$ and $(j,j+1)$ contribute $+1$ and $-1$ to the set discrepancy of $(S_{2k_1-1}^\prime, S_{2k_1}^\prime)$, respectively, or vice versa, before swap removal and addition. The swap $(i,i+1)$ or $(j,j+1)$ that increases the set discrepancy of $(S_{2k_1-1}^\prime, S_{2k_1}^\prime)$ before swap removal and addition and should have been included in $I^*$ (note that this swap does not share an element with the swaps in $I^*$), contradicting the fact that $(i,i+1),(j,j+1)\notin I^*$. Therefore, each added swap contributes $2$ to the total set discrepancy. Moreover, added swaps do not share an element with each other or with other swaps in $I^*$, because two swaps that share an element cannot both contribute $2$ to the total discrepancy.

Therefore, the added $\text{in}(V_{\subseteq})$ swaps contribute $2*\text{in}(V_{\subseteq})$ to the total discrepancy, while the removed swaps reduce the total discrepancy by $2*d(V_{\subseteq})$. Since $\text{in}(V_{\subseteq}) > d(V_{\subseteq})$, this implies that the total set discrepancy increases after swap removal and addition, which contradicts the assumption of maximality of the total set discrepancy indeuced by $I^*$. Hence, $\text{in}(V_{\subseteq})\le d(V_{\subseteq})$.
\end{IEEEproof}
We now prove~\eqref{eq:equivvertexineq} for the case when $|E^j_{\text{swp}}|+1-|V^j_{\text{swp}}|>0$. Suppose to the contrary that \eqref{eq:equivvertexineq} does not hold when $|E^j_{\text{swp}}|+1-|V^j_{\text{swp}}|>0$. Then, $\text{in}(V^j_{\text{swp}})> |E^j_{\text{swp}}|$. By Proposition \ref{prop:removeandadd}, this is impossible since $d(V^j_{\text{swp}})= |E^j_{\text{swp}}|$. 

It remains to prove~\eqref{eq:equivvertexineq} for the scenario that  $|E^j_{\text{swp}}|+1-|V^j_{\text{swp}}|=0$, i.e., to prove that 
\begin{align}\label{eq:nocyclecase}
    \text{in}(V^j_{\text{swp}})-\text{out}(V^j_{\text{swp}})\le |V^j_{\text{swp}}|-4
\end{align}
for $j\in [J]$ and $|E^j_{\text{swp}}|+1-|V^j_{\text{swp}}|=0$. 
Before the proof, we note that the four elements included in the companion sets $(S_{2i-1},$ $S_{2i})$, $i\in[t]$ are of one of the following ``types'':
\begin{enumerate}
    \item Type $1$: $\{a,a+b,a+b+1,a+2b+1\}$ for some $b\ge 1$, where $\{S_{2i-1},S_{2i}\}=\{\{a,a+2b+1\},\{a+b,a+b+1\}\}$;
    \item Type $2$: $\{a,a+b,a+b+c,a+2b+c\}$ for some $b,c>1$, and $a\ge 1$, where $\{S_{2i-1},S_{2i}\}=\{\{a,a+2b+c\},\{a+b,a+b+c\}\}$;
    \item Type $3$: $\{a,a+1,a+1+b,a+2+b\}$ for some $a,b\ge 1$, where $\{S_{2i-1},S_{2i}\}=\{\{a,a+2+b\},\{a+1,a+1+b\}\}$.
\end{enumerate}
To show that the four elements $\{\ell_1,\ell_2,\ell_3,\ell_4\}$ $\in$ $S_{2i-1}\cup S_{2i}$, $i\in[t]$, are of the three types listed above, we order the four elements in as $\ell_1<\ell_2<\ell_3<\ell_4$, for any $i\in[t]$. Note that by the  balance property of $(S_{2i-1},S_{2i})$, we have $\ell_2-\ell_1=\ell_4-\ell_3$. Only one of the following occurs: (1) $\ell_2-\ell_1=1$, which is of Type $3$; (2) $\ell_2-\ell_1>1$ and $\ell_3-\ell_2=1$, which is of Type $1$; (3) $\ell_2-\ell_1>1$ and $\ell_3-\ell_2>1$, which is of Type $2$. 

We note that Type $3$ companion sets can be excluded when $|E^j_{\text{swp}}|+1-|V^j_{\text{swp}}|=0$, i.e., the connected component $G^j_{\text{swp}}$ is acyclic. For any companion sets $(S_{2i-1},S_{2i})$ of Type $3$, it can be verified that at most $2$ swaps from $\{(\ell_1-1,\ell_1),(\ell_2,\ell_2+1),(\ell_3-1,\ell_3),(\ell_4,\ell_4+1)\}$ that share only a single a element with $S_{2i-1}\cup S_{2i}=\{\ell_1,\ell_2,\ell_3,\ell_4\}$ are allowed in $I^*$. This is because no three swaps can each increase the discrepancy of $(S_{2i-1},S_{2i})$ by $1$.
Moreover, one can always use either of the swaps $(a,a+1)$ or $(a+b+1,a+2+b)$, in addition to the swaps between companion sets $(S_{2i_1-1},S_{2i_1})$ and other companion sets, to further increase the set discrepancy of $(S_{2i_1-1},S_{2i_1})$ by $2$. This implies that there should be a self-loop for $v_{i_1}$ in $V^j_{\text{swp}}$ if $(S_{2i_1-1},S_{2i_1})$ is of Type $3$, contradicting the assumption that $G^j_{\text{swp}}$ is acyclic.

We characterize next some properties of companion sets of Type $1$ and Type $2$ in the following proposition. The proof is delegated to the Appendix.
\begin{proposition}\label{prop:type}
For any $i\in[t]$, let $(S_{2i-1},S_{2i})$ companion sets of Type $j$, $j\in[2]$. Then, 
\begin{align}
d(v_i)+d_{\text{out}}(v_i)= j+2.
\end{align}
\end{proposition}
We are now ready to prove~\eqref{eq:nocyclecase}. From Proposition~\ref{prop:removeandadd}, we have $d_{\text{in}}(v_i)\le d(v_i),$ since $d_{\text{in}}(v_i)=\text{in}(v_i)$. From Proposition ~\ref{prop:type}, we have $d_{\text{out}}(v_i)\ge 3-d(v_i)$. Hence,  $d_{\text{in}}(v_i)-d_{\text{out}}(v_i)\le 2d(v_i)-3$. Summing over all $v_i\in V^j_{\text{swp}}$, we obtain
\begin{align*}
    d_{\text{in}}(V^j_{\text{swp}})-d_{\text{out}}(V^j_{\text{swp}})
    &\le 2\sum_{v_i\in V^j_{\text{swp}}}d(v_i)-3|V^j_{\text{swp}}|\\
    &=4|E^j_{\text{swp}}|-3|V^j_{\text{swp}}|\\
    &=|V^j_{\text{swp}}|-4,
\end{align*}
which implies~\eqref{eq:nocyclecase} and therefore completes the proof.

\bibliographystyle{IEEEtran}
\bibliography{biblio1.bib}

\appendix
\section*{Proof of Proposition \ref{prop:type}}
Note that swaps involving elements from Type $1$ companion sets can be grouped into two sets, $\{(a-1,a),(a+b+1,a+b+2),(a+2b,a+2b+1)\}$ and $\{(a,a+1),(a+b-1,a+b),(a+2b+1,a+2b+2)\}$, where swaps in the same group can simultaneously contribute $+1$ to the discrepancy of the companion sets $(S_{2i-1},S_{2i})$, while swaps from different groups cannot simultaneously contribute $+1$ to the set  discrepancy. Hence, $I^*$ cannot include swaps from both groups. When
\begin{align*}
&d(v_i)=|I^*\cap (\{(a-1,a),(a+b+1,a+b+2),\\&(a+2b,a+2b+1)\}\cup\{(a,a+1),(a+b-1,a+b),\\&(a+2b+1,a+2b+2)\} )|=0,    
\end{align*}
i.e., when $v_i$ is an isolated node in $G_{\text{swp}}$, 
one of the groups of swaps corresponds to three outgoing arcs from node $v_i$ in $E_{\text{pot}}$, meaning that $d_{\text{out}}(v_i)=3$. When $d(v_i)\ne 0$, 
without loss of generality, we may assume that 
$$
I^*\cap \{(a,a+1),(a+b-1,a+b),(a+2b+1,a+2b+2)\}= \varnothing.
$$
Then, 
$$
d(v_i)=|I^*\cap \{(a-1,a),(a+b+1,a+b+2),(a+2b,a+2b+1)\}|.
$$
The swaps in $\{(a-1,a),(a+b+1,a+b+2),(a+2b,a+2b+1)\}\backslash I^*$ correspond to $d_{\text{out}}(v_i)=3-d(v_i)$ outgoing arcs from $v_i$ in $E_{\text{pot}}$. In either case, we have $d(v_i)+d_{\text{out}}(v_i)=3$.

Similarly, we group swaps involving elements from Type $2$ companion sets into two sets, $\{(a-1,a),(a+b,a+b+1),(a+b+c,a+b+c+1),(a+2b+c-1,a+2b+c)\}$ and $\{(a,a+1),(a+b-1,a+b),(a+b+c-1,a+b+c),(a+2b+c-1,a+2b+c)\}$. Similar arguments as the one previously described may be used when $d(v_i)+d_{\text{out}}(v_i)=4$. consequently, $d(v_i)+d_{\text{out}}(v_i)=j+2$. 

\end{document}